\documentstyle[aps,prl,preprint,psfig,tighten]{revtex}

\newcommand{\beq}{\begin{equation}}
\newcommand{\eeq}[1]{\label{#1} \end{equation}}
\newcommand{\beqar}{\begin{eqnarray}}
\newcommand{\eeqar}[1]{\label{#1} \end{eqnarray}}

\begin{document}
\draft
\preprint{CU-TP-929}

\title{Anti-Hyperon Enhancement through  Baryon Junction Loops}

\author{Stephen E. Vance$^{1,2}$ and  Miklos Gyulassy${^1}$}
\address{$^1$ Physics Department, Columbia University, 
538 West 120th Street, New York, NY 10027\\
$^2$ Institute for Nuclear Theory, University of Washington, Box 351550, 
Seattle, WA 98195}

\date{\today}

\maketitle

\begin{abstract}
The baryon junction exchange mechanism recently
proposed to explain  valence 
baryon number transport in nuclear collisions
is extended to study midrapidity 
anti-hyperon production.
Baryon junction-antijunction $(J\bar{J})$ loops 
are shown to enhance $\bar{\Lambda},\bar{\Xi},\bar{\Omega}$
as well as lead to long range rapidity correlations.
Results are compared to recent  WA97 $PbPb\rightarrow Y\bar{Y}X$ data.
\end{abstract}
\pacs{25.75.-q,24.10.Lx}

\narrowtext

Striking nonlinear 
nuclear enhancements of multistrange hyperons and anti-hyperons 
have been recently reported for central $Pb+Pb$ collisions at incident 
momentum of 158A GeV \cite{wa97_antib,na49_s98,na49_xi}. 
For example, the  midrapidity density of $\Xi^-(ssd)$
is claimed\cite{na49_xi} to be enhanced by an order
of magnitude relative to the linear $A^{1.0}\sim 200$ extrapolation 
from $pp$ data. Multiparticle production models 
such as HIJING\cite{hijing}, FRITIOF\cite{fritiof}, and 
DPM\cite{dpm_94}, which are based on the
diquark-quark string picture for excited baryons,
underestimate the midrapidity $\Lambda$ yield by a factor of 6 (see 
\cite{topor_95}) and  
the $\Omega^-$ yield by a factor of $10^3$ (see Figure 2).
Recently, diquark breakup mechanisms have been proposed
\cite{kop_89,dpm_dqb,khar_bj96,vance_hijb} 
that can account for at least part of 
the leading ``valence'' hyperon enhancements,
but thus far none of those models
have been able to account for the 
observed $\bar{\Lambda},\bar{\Xi},\bar{\Omega}$ enhancements.  

Large hyperon and anti-hyperon enhancements are of particular interest 
since they have been proposed as one of the signatures of the 
quark-gluon plasma formation in nuclear 
collisions\cite{rafelski_82,rafelski_prl82,koch_86}.
We argue that an equilibrated plasma interpretation of this data 
is problematic since chemical equilibration is slow and since 
adjustable fugacity parameters are needed to reproduce 
the anti-hyperon yields.   In addition, still more parameterizations of 
the baryon chemical potential and transverse collective velocity fields 
are needed in order to fit the detailed rapidity 
and transverse momentum distributions of the hyperons and the anti-hyperons, 
which differ considerably.

Another non-equilibrium mechanism which mimics the effects of an 
equilibrated plasma by being able to enhance the 
strangeness yields is the fusion of strings into 
color ``ropes''\cite{biro_ropes,rqmd}.  
Increasing the string tension or energy density of the string increases 
greatly the Schwinger tunneling probability
\cite{schwinger_51,andersson_83} 
of $s\bar{s}$ and $(ss)(\bar{s}\bar{s})$ production.
However, the main drawback of the rope model 
is that the rope tension or string radius is unconstrained 
by $pp$ and $pA$ data and is taken as a free parameter to fit $AA$ data.
Ropes provide therefore the same freedom to fit data as fugacities,
chemical potentials, and flow velocities in the above plasma
fireball models.

In this letter, we propose instead a novel non-equilibrium hadronic
mechanism derived from the Y-shaped ($SU_c(3)$) gluon structure of 
the baryon to explain these observations.  In addition, this mechanism has
the virtue that its parameters are in principal fixed by $pp$ and 
$pA$ phenomenology.  In particular, we extend the valence baryon junction 
exchange mechanism in \cite{khar_bj96,rossi_77} by including
junction-antijunction $(J\bar{J})$ loops that naturally arise
in Regge phenomenology. 

The concept of a baryon junction, $J(x)$, 
which links the three color flux (Wilson) lines flowing from 
the valence quarks, was developed in the context of 
Regge theory \cite{rossi_77}. 
Several states, such as the quarkless hybrid glueball 
$(J\bar{J}$ or $M^J_0)$ and hybrid exotic mesons 
[$(qJ\bar{J}\bar{q}=M^J_2)$ and $(qqJ\bar{J}\bar{q}\bar{q}=M^J_4)$], 
were proposed by linking junctions, anti-junctions and quarks in 
different ways.  Historically\cite{rossi_77},
the $M^J_0$ trajectory was invoked to explain the difference in 
multiplicities between the total inelastic cross sections 
of $p+\bar{p}$ and $p+p$ scattering.  However, since the experimental 
results were not conclusive (due to the low energies involved) and since
the $M^J_i$ states have not been observed spectroscopically (presumably
due to its large decay width), the junction concept
lay dormant until recently.

A new test of the junction picture of the baryons was 
proposed in \cite{khar_bj96} based on an analysis of the 
valence $(p-\bar{p})$ rapidity distributions in high energy reactions. 
This baryon junction exchange mechanism for the flow of the baryon 
number was implemented in the HIJING/B event generator 
in \cite{vance_hijb}.  It was shown to provide 
not only a very effective mechanism for the transport of {\em valence} 
baryon number over a large rapidity intervals in nuclear collisions 
as observed in \cite{na49_bstop}, but it also was shown to naturally enhance 
the ``valence'' hyperon ($Y-\bar{Y}$) production. 
However, as shown below, this diquark breakup mechanism
is unable to explain the observed anti-hyperon enhancement.

The $M^J_0$ or $J\bar{J}$ trajectory corresponds
in operator form to the exchange of a closed
three string configuration (three sheet topology) written below, 
\beqar
M^J_0  = & \epsilon^{j_1 j_2 j_3} \epsilon_{k_1 k_2 k_3} &
\left [ P \exp \left (ig \int_{x_{\bar{J}}}^{x_J} dx^{\mu} A_{\mu} \right ) 
\right ]^{k_1}_{j_1}  
\left [P \exp \left (ig \int_{x_{\bar{J}}}^{x_J} dx^{\mu} A_{\mu} \right )  
\right ]^{k_2}_{j_2}  \nonumber \\ & \times &
\left [P \exp \left (ig \int_{x_{\bar{J}}}^{x_J} dx^{\mu} A_{\mu} \right ) 
\right ]^{k_3}_{j_3} \;\; .
\eeqar{mj0}
When in a highly excited state, this $J\bar{J}$ three string configuration
fragments into a baryon and an antibaryon
with approximately three times the rapidity 
density of mesons of $e^+e^-$ between the baryon pair. 
The fragmentation of the three independent strings
naturally enhances the transverse momentum as well as the strangeness
content of the baryon and the anti-baryon.
In particular, the fragmentation of $J\bar{J}$ 
into $\Omega^-$ and $\bar{\Omega}^+$ is possible.
In addition, novel long range rapidity correlations are predicted
by this mechanism as shown below.

The generalized optical theorem and Regge theory\cite{collins} provide
the calculus to compute the differential 
inclusive cross section of baryons resulting from $J\bar{J}$ exchange.  
The Regge diagram shown in Fig 1a represents the extra contribution to the 
single inclusive ``valence'' baryon distribution due to the  
junction exchange \cite{khar_bj96,collins} and gives 
\beq
E_B \frac{d^3\sigma}{d^3 {\bf p}_B} 
\rightarrow C_B e^{-(\alpha_{M^J_0}(0)-1)y_B},
\eeq{bj} 
where $C_B$ is an unknown function of the transverse momentum of the
``valence'' baryon and the couplings of the $M^J_0$ Regge state to 
the Pomeron and the baryons.  
Another diagram (same as Fig 1a, but where the $P$ 
and the $M^J_0$ are interchanged) gives rise to the contribution 
of $\exp((\alpha_{M^J_0}(0)-1)y_B)$ for the target region.
The contribution to the double differential inclusive cross section 
for the inclusive production of a baryon and an anti-baryon  
in $N N$ collisions due to $J\bar{J}$ exchange (shown diagrammatically 
in Fig 1b) is
\beq
E_B E_{\bar{B}} \frac{d^6\sigma}{d^3 {\bf p}_B d^3 {\bf p}_{\bar{B}}} 
\rightarrow C_{B\bar{B}} e^{(\alpha_{M^J_0}(0) - 1)
|y_B - y_{\bar{B}}|}, 
\eeq{bantib}
where $C_{B\bar{B}}$ is an unknown function of the transverse momentum and 
the $M^J_0+P+B$ couplings.
We include these processes in the string model by 
introducing diquark breaking (Fig. 1c) and $J\bar{J}$ loop (Fig.1d)
string configurations.  

From Eq. (\ref{bantib}), the predicted rapidity correlation 
length $(1-\alpha_{M^J_0}(0))^{-1}$ depends upon the value of the 
intercept $\alpha_{M^J_0}(0)$.
As in \cite{vance_hijb}, we assume $\alpha_{M^J_0}(0) 
\simeq 1/2$.   Our choice is based upon arguments and estimates from a 
multiperipheral model approximation\cite{rossi_77,eylon_74} 
and is consistent with the multiplicity and energy 
dependence of $\Delta \sigma = \sigma_{p\bar{p}} - \sigma_{pp}$
\cite{svance_alpha}.    This value $\alpha_{M^J_0}(0)$ can now
again be tested through the rapidity correlations.
To test for this $M_J^0$ component ($\alpha_{M^J_0}(0) \simeq 1/2$)
requires the measurement of rapidity correlations on a scale
$|y_B-y_{\bar{B}}| \sim 2$.  
We note that it would be especially useful to look for
this correlation in high energy $p+p \rightarrow B + \bar{B} + X$,
especially at RHIC where very high statistics multiparticle
data will soon be available.

Replacing the $M^J_0$ reggeon in Fig 3b with the Pomeron or other 
Reggeon states gives the remaining contributions to the double 
inclusive differential cross section.    Replacing the 
$M^J_0$ with the Pomeron leads to a uniform rapidity distribution 
which corresponds to the uncorrelated product
of two single inclusive differential cross sections\cite{collins}.
(The short range $\Delta y\sim 1$ dynamical (Schwinger) 
correlations in string models lie outside the Regge kinematic region).  
Replacing the $M^J_0$ with other Reggeons such as the $\omega$
provides a contribution which is similar to Eq. (\ref{bantib}) but 
with a different coupling and with 
$\alpha_R(0)$ instead of $\alpha_{M^J_0}(0)$.  We argue that 
the contributions from these Reggeons is small since the
contribution from the $M^J_0$ seems to dominate the contribution 
from the $\omega$ in $pp$ and $p\bar{p}$ scattering
(see \cite{rossi_77,svance_alpha} and references therein).    
In addition, in the string picture, the contribution of the 
$\omega$ is modeled as single string between the baryon and the 
antibaryon which leads to {\underline {lower}} multiplicities 
and does {\em not} provide a strangeness
enhancement.

We modified HIJING/B\cite{vance_hijb} to include the 
above $J\bar{J}$ loops and the new version, 
called HIJING/${\rm B\bar{B}}$, is available at \cite{hjbbar}.
By fitting $\bar{p}$ and $\bar{\Lambda}$ data
from $p+p$\cite{pp400} and $p+S$\cite{na35_hyperons,na35_charged}
interactions at 400 GeV/c and 200 GeV/c incident momentum, respectively, 
the cross section for $J\bar{J}$ exchange is found to be
 $\sigma_{B\bar{B}} = 6$ mb.  We took the threshold 
cutoff mass of Fig. 1d configurations to be
$m_c = 6$ GeV.  This large minimum cutoff is necessary in order 
to provide sufficient kinematical phase space for fragmentation of the
strings and for $B\bar{B}$ production.
In addition, the applicability of the Regge analysis also requires 
a high mass between all of the external lines in Fig 1b.  
At SPS energies ($\sqrt{s}\sim 20$), this kinematic constraint
severely limits the number of $J\bar{J}$ configurations allowed, 
reducing its effective cross section to $\sim 3$ mb.
The $\vec{p}_T$ of the baryons from the $J\bar{J}$ loops 
are obtained by adding the $\vec{p}_T$ of the three sea quarks 
along with an additional soft $p_T$ kick.
The soft $p_T$ kick is taken from a Gaussian distribution whose 
width $\sigma = 0.6 \;{\rm GeV/c}$ is fit to $p+S$ 
data\cite{na35_hyperons} at 200 GeV/c.
At collider energies, hard pQCD processes are modeled as kinks in strings 
as in HIJING\cite{hijing}.

In multiple collisions, a (color 6)
diquark broken through a junction exchange
in a previous collision has a finite probability
to reform (color $\bar{3}$) in a subsequent collision.  
However, in the present implementation, we assume that when a junction
loop is formed in a previous collision, then it cannot be destroyed in
subsequent collisions.  The $pA$ data are 
consistent with these assumptions.   We emphasize that the present 
version of HIJING/${\rm B\bar{B}}$ does not include final state 
interactions.    Although final state interactions 
such as (e.g. $N + \pi \rightarrow \Lambda + K$) 
and $Y + \bar{p} \rightarrow X$ can change the flavor and 
relative yields of many hadrons, we note that these interactions 
cannot easily enhance the multiplicities of the multi-strange baryons, $\Xi$
and $\Omega$.  

In Figure 2, we show the computed net hyperon yields $(Y+\bar{Y})$
from the HIJING, HIJING/B and HIJING/${\rm B\bar{B}}$ event generators for 
$p+Pb$, $S+S$ and $Pb+Pb$ at 160 AGeV.  HIJING (open triangle)
is seen to underpredict the hyperons in $Pb+Pb$ reactions.
HIJING/B results (open squares) show that the valence junction exchange 
significantly enhances the multiple strange baryon yields.   
The anomalously large WA97 $\Omega$ yield (solid triangle)
in $PbPb$ is not explained however.
The $J\bar{J}$ mechanism in HIJING/${\rm B\bar{B}}$
enhances the net $(Y+\bar{Y})$ yield only modestly.
However, Figure 3 shows the large effect of the $J\bar{J}$ loops
on the ratio of the yields of the antihyperons to the hyperons, 
$\eta = N_{B}/N_{\bar{B}}$.
In HIJING, this ratio is close to 1 for $S = -2, -3$ baryons because
they are mostly formed via diquark-anti-diquark fragmentation of single
strings.   In HIJING/B, the splitting of the string in Figure 1c into 
three smaller mass strings strongly reduces the phase space for 
antihyperon production and leads to a very small ratio of $\eta$ 
shown in Fig. 3.  HIJING/${\rm B\bar{B}}$ solves this problem by providing
a natural channel that enhances anti-hyperons production.

Returning to the striking enhancement of the $\Omega$ in Figure 2,
we note that the  diamonds show that even a very modest rope 
enhancement of $\kappa = 1.4 \; {\rm GeV/fm}$ is enough to 
enhance the $\Omega$ by
another factor of 10.  This rare observable is exponentially
sensitive to uncertainties in the multiparticle phenomenology.
We note for example that there exist more complex
multibaryon loop diagrams of junction  - anti-junction vertices 
that can further enhance $\Omega\bar{\Omega}$ production.  
We do not pursue such possibilities here, but leave them for another
work, since our main point here is to show that $J\bar{J}$ loops 
are natural and provide a qualitative explanation of the observed 
moderate antihyperon/hyperon ratios in Fig.3.

Finally, we predict in Fig. 4 the initial distribution of antibaryons 
for $Au+Au$ collisions at $\sqrt{s} = 200A$ GeV for $b\le 3 \;{\rm
fm}$.  The estimates for the $\bar{p}$ in HIJING/${\rm B\bar{B}}$ are 
a factor of 3 lower than the original HIJING prediction.  
This results from correcting for an overpredicted (factor of 2)
$\bar{p}$ yield by HIJING for $p+p$ collisions at 400 GeV/c and allowing
at most one  $J\bar{J}$ loop in the wounded nucleon string.
The $\bar{\Lambda}$ are less effected (factor of 2.5) by these corrections
due to the extra strangeness enhancment from the fragmentation of 
the $J\bar{J}$ loops.

In summary, we showed that modifications in the multiparticle production 
dynamics which are consistent with $pp$ and $pA$ physics can yield
significant (anti)hyperon production as 
well baryon stopping power in $AA$ collisions.
The large WA97 antihyperon/hyperon yields require $J\bar{J}$ loops in 
this approach.  However, the absolute yields of
especially the $\Omega^-(sss)$ are still underestimated (without ropes
or more complex multibaryon junction diagrams). 
Upcoming measurements at RHIC on baryon number transport and
$B\bar{B}$ correlations will be especially important in clarifying 
whether indeed the $J\bar{J}$ exchange mechanism 
plays a dominant role in hyperon dynamics.

This work was supported by the Director, Office of Energy Research,
Division of Nuclear Physics of the Office of High Energy and
Nuclear Physics of the U.S. Department of Energy under Contract No.
DE-FG02-93ER40764. S.E. Vance also thanks the Institute for Nuclear Theory at
the University of Washington for its hospitality during which part of 
this project was completed.

\newpage
\begin{figure}[htb]
\hspace{1.0in}

\psfig{figure=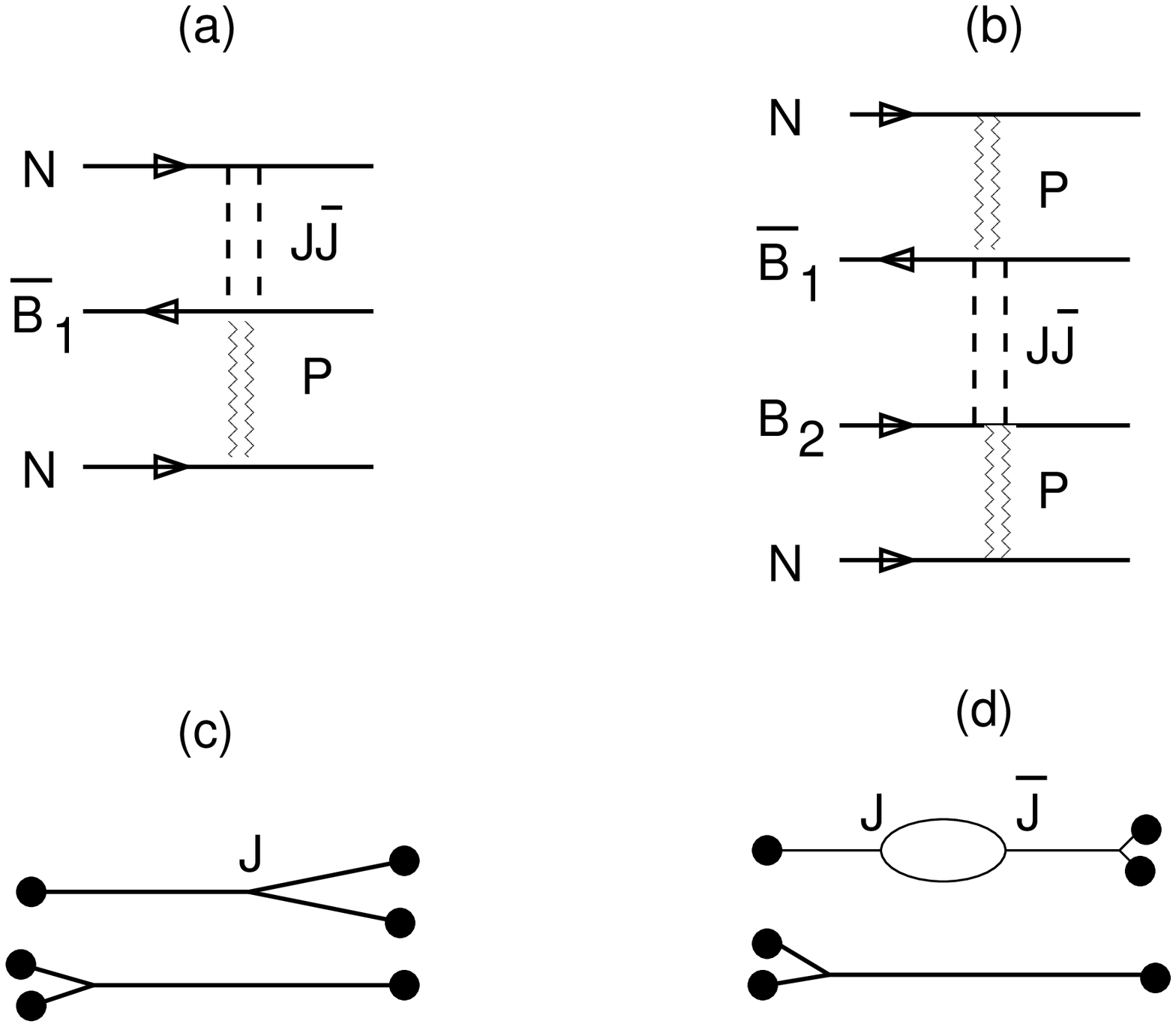,height=6.0in,width=6.0in,angle=0}
\caption{The Regge diagrams for the single baryon junction exchange 
and $J\bar{J}$ loops are shown in (a) and (b), respectively.  The string
model implementation of each Regge diagram are shown in (c) and (d). }
\end{figure}

\begin{figure}[htb]
\vspace{3cm}
\hspace{1.0in}
\psfig{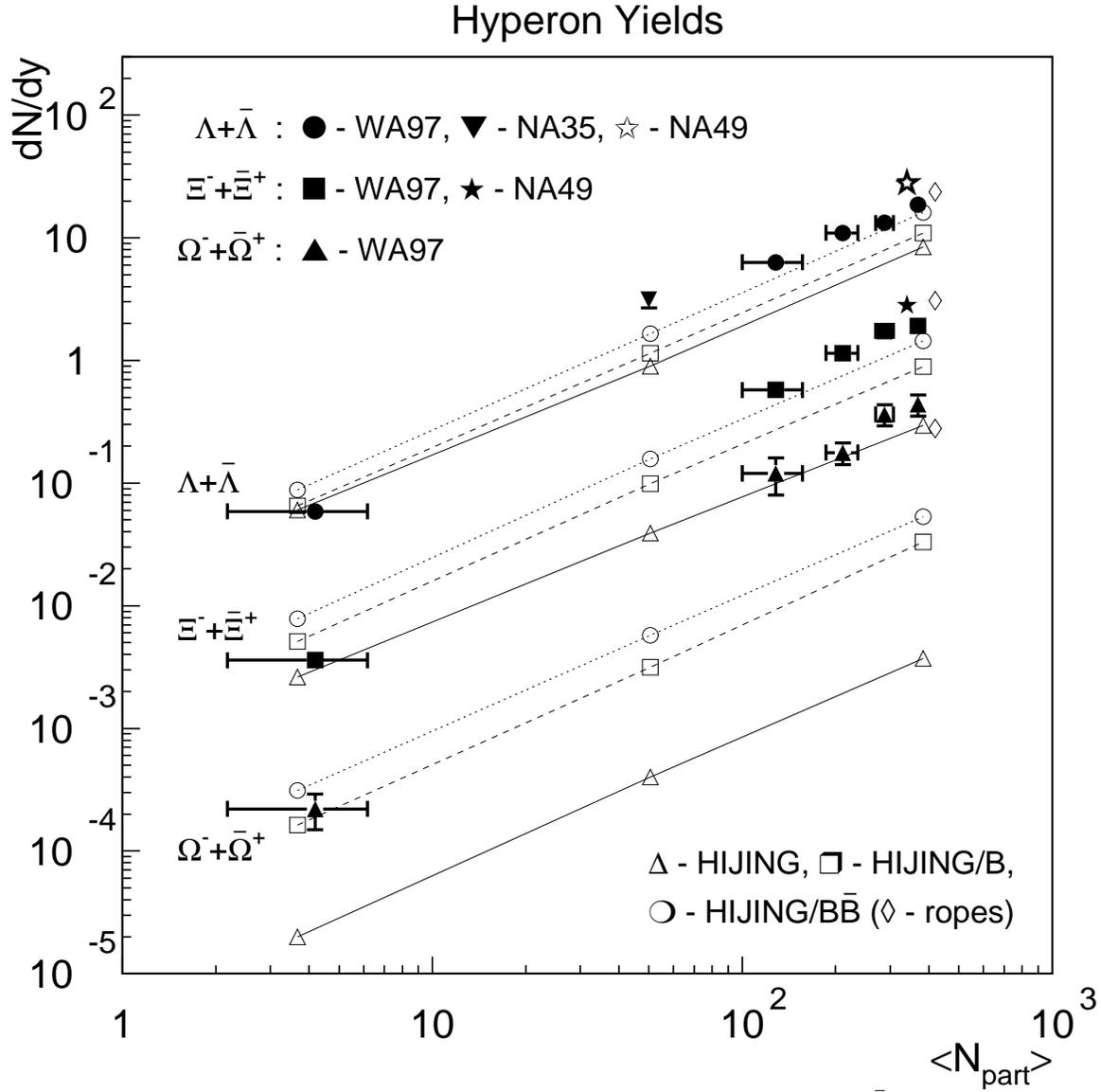}
\caption{Hyperon yields from HIJING, HIJING/B and HIJING/$B\bar{B}$ 
for $p+Pb$, $S+S$ and $Pb+Pb$ at incident momentum 
$p_{lab} = 160$ AGeV are shown
along with data from the NA35 \protect \cite{na35_hyperons}, the 
NA49 \protect \cite{na49_s98,na49_xi} and the 
WA97 \protect \cite{wa97_antib} collaborations.}
\end{figure}

\newpage
\begin{figure}[htb]
\vspace{3cm}
\hspace{1.0in}
\psfig{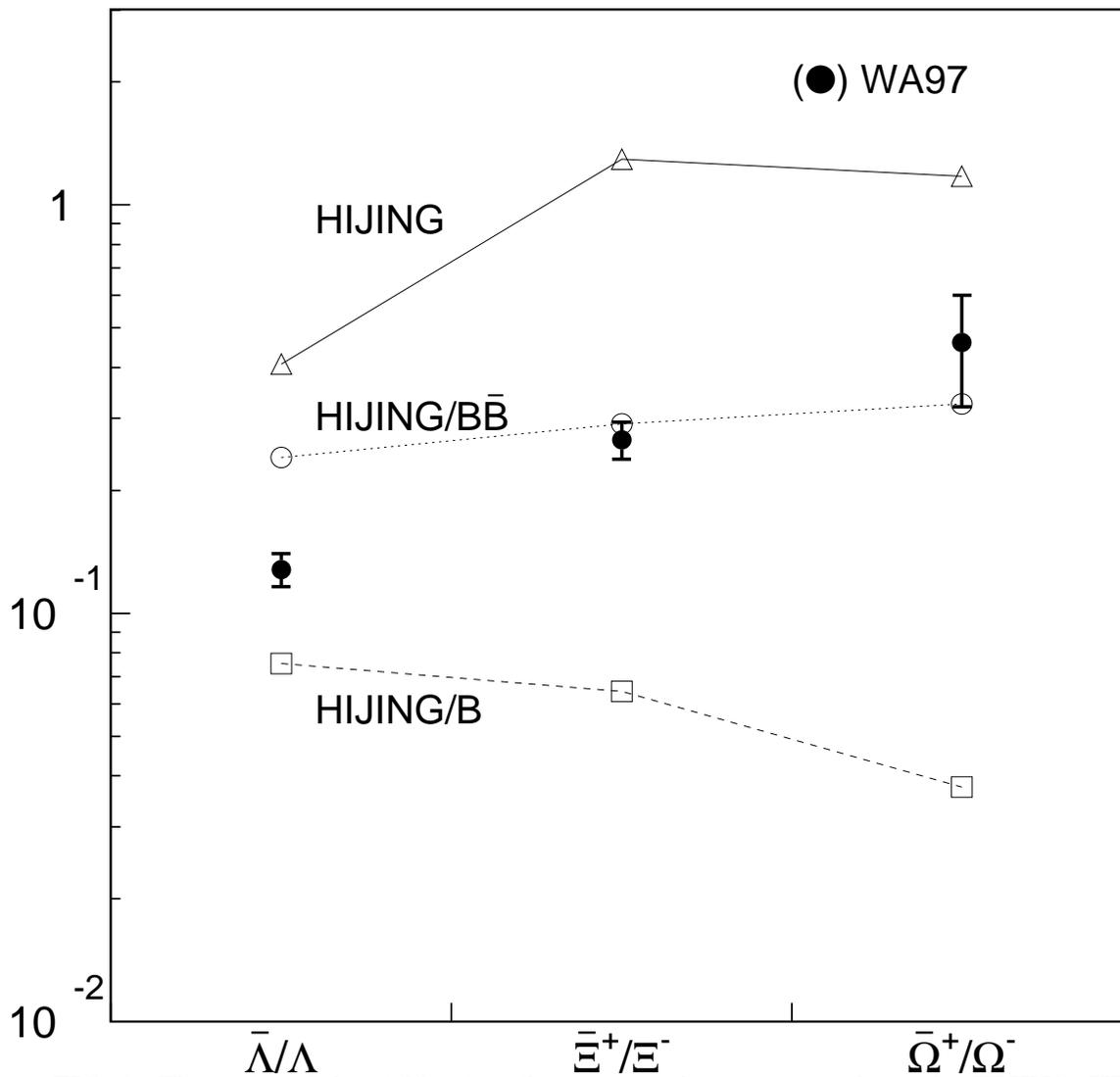}
\caption{The ratios of the yields of antihyperons to hyperons are shown
for HIJING, HIJING/B and HIJING/$B\bar{B}$ for $p+Pb$, 
$S+S$ and $Pb+Pb$ at incident momentum $p_{lab} = 160$ AGeV along with 
data from the WA97 \protect \cite{wa97_antib} collaboration.}
\end{figure}

\newpage
\begin{figure}[htb]
\vspace{3cm}
\hspace{1.0in}
\psfig{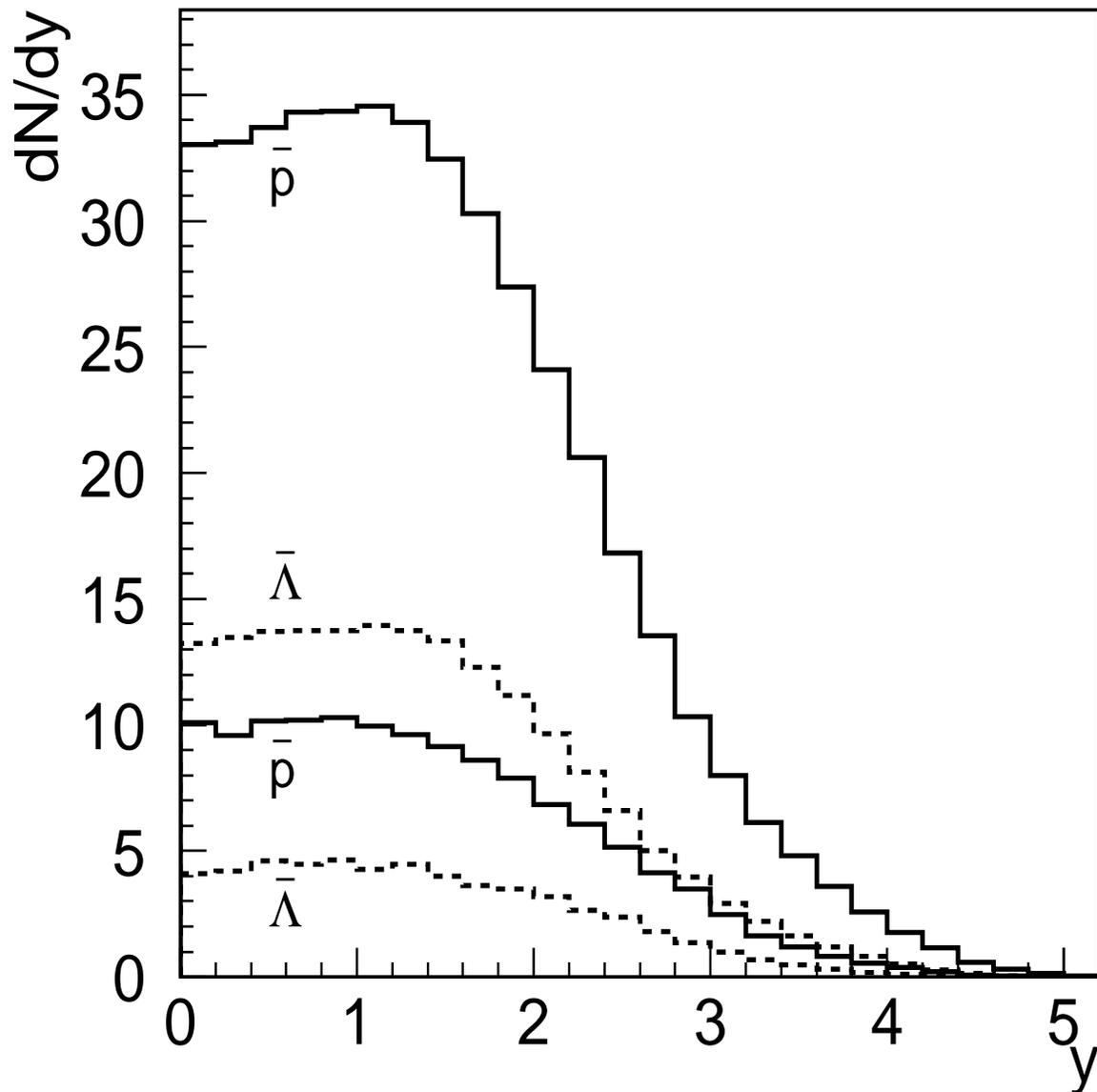}
\caption{Predictions for the initial $\bar{p}$ (solid lines) 
and $\bar{\Lambda}$ (dashed lines)
rapidity distributions are given for HIJING (upper solid and dashed curves) 
and for HIJING/${\rm B\bar{B}}$ (lower solid and dashed curves) for 
Au+Au collisions at $E_{cm} = 200$ at $b \le 3 \;{\rm fm }$.}
\end{figure}

\end{document}